# On the significance of elastic coupling for stresses and leakage in frictional contacts

Christian Muller[1], Martin H. Muser[1,2], Giuseppe Carbone[3], and Nicola Menga[3*]

[1]*INM - Leibniz Institute for New Materials, 66123 Saarbrücken, Germany*
[2]*Dept. of Materials Science and Engineering, Saarland University, 66123 Saarbrücken, Germany and*
[3]*Dept. of Mechanics, Mathematics, and Management, Politecnico di Bari, 70125 Bari, Italy*

We study how the commonly neglected coupling of normal and in-plane elastic response affects tribological properties when Hertzian or randomly rough indenters slide past an elastic body. Compressibility-induced coupling is found to substantially increase maximum tensile stresses, which cause materials to fail, and to decrease friction such that Amontons' law is violated macroscopically even when it holds microscopically. Confinement-induced coupling increases friction and enlarges domains of high tension. Moreover, both types of coupling affect the gap topography and thereby leakage. Thus, coupling can be much more than a minor perturbation of a mechanical contact.

Explaining and predicting the properties of interfaces between solid bodies requires a proper description of mechanical contacts at microscopic scales. This is because small-scale roughness, which is even present on nominally flat solids, makes true contact be smaller, often distinctly smaller than if the contacting surfaces were atomically flat [1–3]. The zones of non-contact cause interfacial electric- and heat-flow resistances [4, 5] as well as extra mechanical compliances in normal and tangential direction [6, 7]. Adhesion can be strongly reduced [8–10] and fluid may leak through the thin gap between a surface and a seal [11, 12]. To assess structural and mechanical properties of contacts—most notably, contact area, stress, and gap distributions—in the important limiting case of linearly (visco-) elastic solids, the in-plane and out-of-plane elastic coupling, in the following simply refered to as *coupling*, is commonly neglected [13]. Yet, shear stresses acting on originally flat surfaces do induce normal displacements or stresses already in linear order, unless the solid is semi-infinite and incompressible. Without coupling, the sound of friction, be it caused by violins or squeeling breaks [14], would often be different and Schallamach waves [15, 16], which are kinetic-friction induced buckling instabilities of elastomers, would not be possible. Generally speaking, coupling has a destabilizing effect on sliding friction and weakens frictional cracks [17].

While a description of the just-mentioned phenomena requires approaches beyond either linear elasticity or quasi-static conditions, quite a few questions related to linear coupling have not yet found satisfactory answers even for steady-state sliding. For instance, how does friction affect size and shape of the true contact during sliding [18, 19], do moving seals seal better than static seals, and, given a microscopic friction coefficient for planar surfaces, does roughness increase or decrease the macroscopically measured friction? How do changes compare to those induced by loading configuration in soft-matter systems, which were reported to be of order $\mathcal{O}(10-30\%)$ and sometimes substantially more [20, 21]?

In this Letter, we explore these and related questions for various rigid indenters sliding past an elastic solid with arbitrary contact modulus $E^*$, Poisson's ratio $\nu$, and height $h$ (see Fig. 1(a)), assuming steady-state conditions and Amontons' microscopic friction. Although some of the issues raised here have already been addressed in line contacts [22–24], load-area and other relations do not generalize from line to areal contacts, neither in simple indenter geometries [25] nor in randomly rough contacts [26, 27], so that the effect of roughness on the friction coefficient can differ between the two cases. More importantly, the analysis of how coupling affects leakage cannot be addressed in line contacts, since they automatically seal in the lateral (sliding) direction, while they are open in the transverse direction. In contrast, percolation of randomly rough two-dimensional surfaces is isotropic, even if roughness and flow factors are not [28]. Moreover, still unexplored is the important effect of linear coupling on either von Mises or maximum tensile stresses in rough contacts, although they are crucial for the onset of plastic deformation and the mechanical failure of materials, respectively.

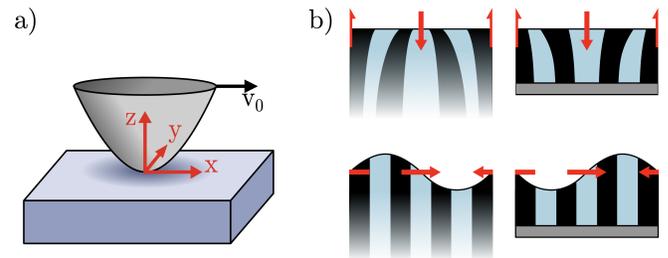

FIG. 1. (a) Contact set-up: a rigid indenter sliding along the $x$ axis at constant velocity $v_0$. Vector components in $x$, $y$, and $z$ direction are called longitudinal, transverse, and normal, respectively. (b) Cross-sections of a compressible (left) and confined (right) body loaded by sinusoidal surface stesses, whose extrema are indicated by red arrows. Stripes and shapes represent coupling-induced longitudinal (top) and normal (bottom) displacements, respectively.

We solve the contact problem numerically using Green's function molecular dynamics (GFMD) [29],

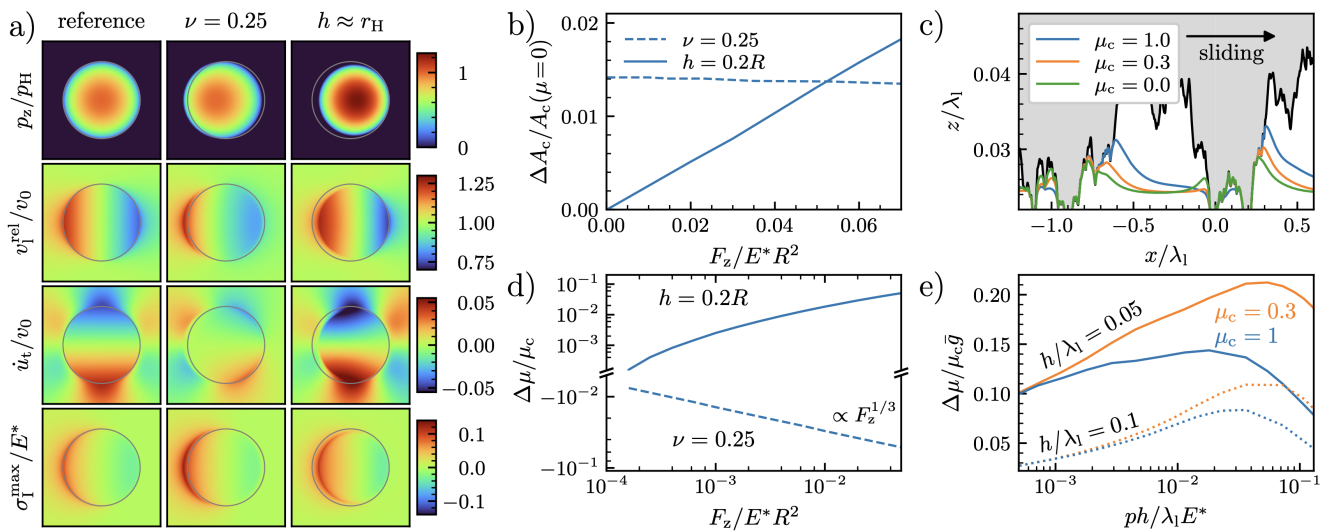

FIG. 2. Panels (a,b,d) and (c,e) relate to Hertzian indenters of radius R and randomly rough indenters, respectively. (a) Contact pressure $p_z$ (top row) normalized to the maximum regular Hertzian contact pressure $p_H$, relative longitudinal $v_l^{rel}$ (second row), transverse $\dot{u}_t$ (third row) velocity, and the maximum principal surface stress $\sigma_I^{max}$ (bottom row) for the reference ($\nu = 0.49$, $h \to \infty$, left column), the semi-infinite compressible (middle column), and the confined nearly incompressible (right column) elastomer. The applied normal load is $F_z = 0.01\, E^* R^2$ yielding a Hertzian contact radius of $r_H \approx 0.2\, R$ (gray circles). Coupling-induced relative changes in (b) contact area $\Delta A_c / A_c$ and (d) friction coefficient $\Delta\mu/\mu_c$ as functions of dimensionless normal load $F_z / E^* R^2$. (c) Rough contact cross section, and (e) normalized relative change of the friction coefficient vs. the dimensionless normal pressure $ph/E^*\lambda_l$ for confinement-induced coupling. Moreover, $\bar{g}$ and $\lambda_l$ are the rough surface's root-mean-square gradient and long-wavelength cutoff, respectively.

which is a Fourier-based boundary value method to calculate elastic surface displacements under periodic boundary conditions. To elucidate the effects of coupling, the continuum description of the normal displacement in Sect. 2.2.1 of Ref. [30] was generalized to compute the full three-dimensional stress tensor and displacement field for a solid with arbitrary thickness and compressibility. The needed analytical (inverse) Green's functions [23, 31] are summarized in the Supplemental Material [32], along with model and methods details, including the topography generation for rigid, randomly rough indenters and the procedure for the leakage calculation [33]. The essence of coupling effects is depicted in Fig. 1(b). Codes, input files, and results are available [34].

The default value for the microscopic friction coefficient is set to $\mu_c = 1$. A Poisson's ratio of $\nu = 0.25$ is used to analyze the generic behavior of compressible materials and $\nu = 0.49$ for (nearly) incompressible ones. The first value is half way between that of many metals with $\nu \gtrsim 0.3$ and that of many ceramics with $\nu \lesssim 0.2$.

To set the stage for this work, we first establish in Fig. 2 that coupling affects areal and line contacts in a similar fashion [23, 24]. For example, for a Hertzian geometry, Fig. 2(a) reveals that coupling destroys the normal pressure symmetry also in areal contacts, which now entails non-circular contact shapes. More specifically, confinement-induced coupling skews the pressure to the leading edge so that it carries more load than the trailing edge, as in line contacts [24, 35]. This effect can be deduced directly from Fig. 1(b) showing that the shear stress in the center—pointing to the right as does the shear force in Fig. 1(a)—makes the displacement field "want" to lift up near the leading edge, which, to keep the normal load constant and steady sliding conditions, the indenter must counteract with an increased constraint force, in a similar fashion as in viscoelastic contacts [36–38]. By virtue of what could be called a downhill-slope force [39], the leading edge opposes sliding more than the trailing edge pushes the indenter forward, which increases the global friction coefficient from the microscopic value $\mu_c$ to $\mu_c + \Delta\mu$. These trends reverse for compressible elastomers, for which the pressure maximum shifts to the trailing edge, resulting in smaller friction. Changes in global friction can also be related to an interplay of coupling-induced loss of (anti-) symmetry in velocity and stress fields, which alters the local heat production. This argument is presented in detail in the SM (Sec. 2b) together with a compilation of line-scans for displacement, velocity, and stress fields (Fig. SM-2).

In the case of compressibility-induced coupling, the increase of a normal constraining pressure on the trailing edge (preventing surfaces from interpenetrating) and the decrease of the compressive stress on the leading edge can only be proportional to the original normal stress. Ultimately, this is because the coupling correction, as described by the variable $\Phi_{13}$ introduced in the SM, only depends on $q_x/q$ but not explicitly on $q$ when the elastomer is semi-infinite. Consequently, the correction to the downhill force and thus to the friction coefficient is proportional to the mean absolute slope, which, in the case of Hertzian indenters, is proportional to the contact radius $r_c$ and thus to $F^{1/3}$, as is confirmed in Fig. 2(d). Algebraic scaling relations for the confinement-induced coupling do not arise, because the amplitudes of coupling terms have a non-algebraic dependence on film thickness and wave number.

Simulations similar to those for Hertzian indenters were repeated for randomly rough indenters, a cross-section of which is depicted in Fig. 2(c) along with an elastic, confined counterbody. The characteristics of its displacement fields resemble those of Hertzian indenters; however, Fig. 2(e) reveals that the friction coefficient now increases only initially with load before it starts to decrease at high loads. When coupling is caused by finite compressibility, the friction coefficient of a randomly rough indenter is also non-monotonic in load, but trends are inverted again, i.e., it is always lower than $\mu_c$ but increases after the initial decrease with load (not shown explicitly). The predominant reason for the non-monotonicity is that the normal displacement gradients at trailing and leading edges first increase with load, as in a Hertzian geometry, but eventually become smaller with increasing contact dimension due to sinusoidal characteristics of the roughness profile, thereby reducing the down-hill-slope effect.

In addition to friction, von Mises and tensile stresses are central tribological quantities, since they affect the failure of materials. Roughly speaking, ductile solids deform plastically first near defects where the von Mises stress, which is $\sqrt{3/2}$ times the standard deviation of the stress-tensor eigenvalues, is largest, while polymers and brittle materials like ceramics break near points of high tension, given by the largest stress-tensor eigenvalue. So far, analytical solutions for stress distribution below frictional Hertzian indenters have been obtained neglecting coupling [40, 41], in which case the maxima of tensile and von Mises stresses are located in the surface at the trailing edge for $\mu_c > 0.3$, as depicted in Fig. SM-2, row six, second column. Experiments on thin elastomers confirm these trends for line contacts with thin elastomers [35]. For this reason, and because crack initiation is most effective in the near-surface region [42], we focus on surface stresses in the following discussion of coupling effects

Analysis of the stress profiles—details are shown in Fig. SM-2, row 6 being most relevant to this paragraph—reveals a remarkable 30% increase in the tensile stress due to coupling for $\nu = 0.25$ and $r_H/R = 0.2$, whereas the von Mises stress is barely affected. Changes in the stress due to confinement-induced coupling are more difficult to evaluate, because confinement reduces the contact area at given normal load so that the semi-infinite, incompressible elastomer is no longer a good reference. The reduced contact area leads to a dramatic increase of $\sigma_{vM}$. In addition, the zone where the tensile stress is close to its maximum value increases substantially in size. Thus, both types of coupling can strongly enhance the likelihood of crack formation. Of course, linear elasticity can only be used to estimate the onset of plasticity and/or material failure. Once triggered, additional phenomena, which are likely dissipative in nature, occur, thereby altering friction further.

To highlight the importance of in-plane stress and deformations, we redefine the reference to which numerical results are compared. To this end, we first conduct a regular contact-mechanics calculation for a frictionless interface and then add the interfacial shear stress in post-analysis as a perturbation under the assumption that all material points at the interface have the same relative in-plane velocity $v_0$. The top row of Fig. 3 reveals that this procedure substantially underestimates tension. One effect missing in the pursued approach is symmetry breaking, which makes maximum tensile stresses move to the trailing edge for both couplings. Besides this qualitative effect, quantitative differences between the true tensile stress and the one obtained in post-analysis are factors easily surpassing two to four in the studied systems.

Coupling does not only alter stresses but also displacements and thereby the interfacial separation, which will be called gap $g$ in the following. The gap determines the local resistance $\rho$ to in-plane fluid flow in between a rigid surface and a seal. In the Reynolds thin film equation, $\rho \propto 1/g^3$ [12 and 43]. The bottom row of Fig. 3 shows the fluid current for our confined elastomer in four cases, i.e., for two sliding and two flow directions. Leaking matters in the sliding direction for applications like scrubbers and syringes and in the orthogonal direction for rotary seals and journal bearings. The shown images reveal that fluid flow is affected by the sliding direction and thus by coupling.

The flow patterns depicted in Fig. 3 are in line with the idea that fluid-flow is impeded predominantly by a few constrictions [43, 44], where current densities are high and which become critical just before they block fluid flow completely. The average effect of coupling on the fluid flow suffers from large statistical uncertainties near the percolation threshold, since the number of relevant constrictions per unit area is minuscule. Therefore, we studied an individual constriction to isolate the effect of coupling on it. For this purpose, we choose a roughness of square-lattice symmetry having the form $h(\mathbf{r}) = h_0\{\cos(qx) + \cos(qy)\}$, because its relative contact area at the percolation threshold, $a_c^* \approx 0.405$ [33] is close to that of randomly rough surfaces, $a_c^* \approx 0.42$ [12], while the exponent $\zeta = 69/20$ with which the total current disappears near the percolation threshold, $I \propto (a_c^* - a_c)^\zeta$ was found to be identical for square [33] and random roughness in the case of semi-infinite elastic bodies [45].

Using $q = 2\pi/\lambda$, where the wavelength $\lambda$ coincides with the linear dimension of the periodically repeated simulation cell, and $h_0 = \lambda/(2\pi)^2$, we obtain a radius of curvature at the peaks of $R = \lambda$. The indenter is then squeezed against either a semi-infinite elastomer with $\nu = 0.25$ or a $\nu = 0.49$ elastomer slab of height $h = \lambda/10$. In the first case, $a_c^*$ does not change in a frictionless reference, while it increases to essentially $a_c^* = 0.5$ for the confined elastomer. This "canonical percolation threshold" is the one that applies to square or random roughness if points above the mean height are (assumed to be) in contact, while those below it are not. Due to coupling, the contact area increases in both cases compared to the frictionless reference at fixed pressure, i.e., from 0.405 to 0.416 for the $\nu = 0.25$ elastomer and from 0.5 to 0.556 for the confined elastomer. In addition, flow factors are enhanced parallel to the sliding direction in an isolated constriction but blocked in the transverse direction, as depicted in Fig. 3(j-m). Thus, at loads slightly smaller than those needed to reach the percolation threshold in a random surface, roughly half of the critical constrictions would start to block fluid due to friction and coupling, while some previously closed constrictions would open up.

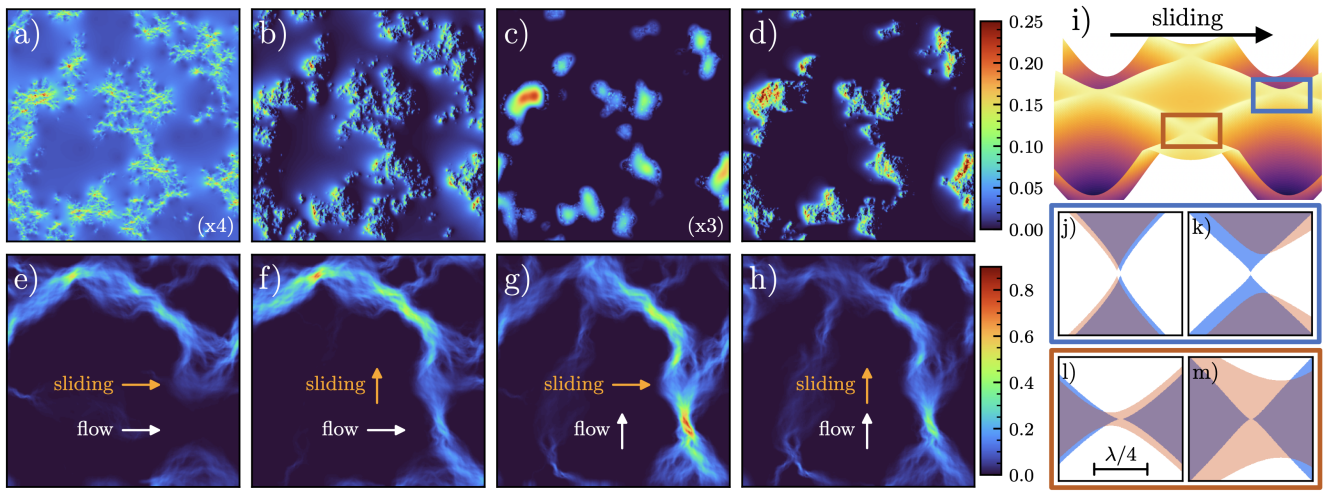

FIG. 3. Top row: maximum tensile stress normalized to $E^*$ for a (a,b) compressible and (c,d) confined elastomer. (a,c) were deduced from static simulations and lateral stress added in post-processing, whereas (b,d) are based on full sliding simulations. Bottom row: (e-h) leakage current *density* for different fluid flow and sliding directions for the confined layer at a relative contact area of approximately 19%. Each panel was produced using the same randomly rough indenter and, for (e-h), the same fluid-pressure difference . Results are normalized to the maximum value in frictionless conditions. (i) 3D contact around the critical constriction for the square roughness profile against either (j,l) a semi-infinite, $\nu = 0.25$ elastomer or (k,m) a confined, $\nu = 0.49$ elastomer. Blue color indicates the frictionless contact area, while orange marks the case with friction. The purple region is the overlap of the two. Constrictions open in sliding direction (j,k) but close in the transverse direction (l,m).

The implications that results in isolated constrictions have for flow in randomly rough contacts cannot be easily ascertained: while the opening of channels in parallel direction facilitates fluid flow in that direction, the percolation is isotropic in the thermodynamic limit [28] so that the blocking of previously open channels in the transverse direction can prevent complete percolation. In fact, after averaging results over eight independent surface realizations using Bruggeman's self-consistent equation [12, 28], we find a reduction of fluid flow in both directions in all cases. For compressibility coupling, under given normal load, it turned out to be 11% and 20% in longitudinal and transverse direction, respectively. For confined elastomers these numbers changed to 27% and 30%. All numbers apply to a relative contact area of $a_c \approx 0.2$, which is still far from the percolation threshold, i.e., relative corrections will be much enhanced as the pressure is further increased.

The simulations presented in this paper reveal that linear coupling can strongly affect all central tribological properties by 10% and even much beyond when the materials in contact are either sufficiently thin or compressible. Specifically, we find that coupling counteracts the validity of Amontons' law and increases both contact area and in-plane tensile stresses at fixed normal load compared to the uncoupled case. Leakage is also impacted by coupling-induced changes in the gap distribution and contact stiffness, entailing an overall reduction of the fluid flow and shifting the percolation threshold to smaller nominal pressures. Thus, the common practice of neglecting coupling can indeed lead to substantial errors in the prediction of interfacial properties as foreseen by Johnson [46].

Seeking for experimental confirmation of our findings may be a challenging task, because nonlinear elasticity is likely to play a role in sliding contact tests [47]. Some studies [35, 48], however, seem to have successfully avoided nonlinear phenomena (e.g. contact shrinking), which makes us optimistic that our findings can be experimentally verified after all. We expect them to matter in a broad variety of systems, i.e., for any rough solid with a Poisson's ratio clearly different from 0.5 or any system of finite thickness, including coatings and confined elastomers. Particular examples would be MEMS, hard antiwear/antifriction or other protective coatings, e.g., on photovoltaic panels but also thin static seals or electric brushes used in sliding electrodes.

---

* nicola.menga@poliba.it

[1] F. P. Bowden, D. Tabor, and F. Palmer. *The Friction and Lubrication of Solids*. Clarendon Press, 1950.
[2] B. N. Persson. Theory of rubber friction and contact mechanics. *Journal of Chemical Physics*, 115:3840–3861, 2001.
[3] T. D. Jacobs and L. Pastewka. Surface topography as a material parameter. *MRS Bulletin*, 47:1205–1210, 12 2022.
[4] J. R. Barber. Bounds on the electrical resistance between contacting elastic rough bodies. *Proceedings of the Royal Society A: Mathematical, Physical and Engineering Sciences*, 459:53–66, 2003.
[5] B. N. Persson. On the electric contact resistance. *Tribology Letters*, 70:1–9, 2022.
[6] A. Baltazar, S. I. Rokhlin, and C. Pecorari. On the relationship between ultrasonic and micromechanical properties of contacting rough surfaces. *Journal of the Mechanics and Physics of Solids*, 50:1397–1416, 2002.
[7] C. Campañá, B. N. Persson, and M. H. Müser. Transverse and normal interfacial stiffness of solids with randomly rough surfaces. *Journal of Physics Condensed Matter*, 23, 2011.
[8] L. Pastewka and M. O. Robbins. Contact between rough surfaces and a criterion for macroscopic adhesion. *Proceedings of the National Academy of Sciences of the United States of America*, 111:3298–3303, 2014.
[9] B. N. Persson and M. Scaraggi. Theory of adhesion: Role of surface roughness. *Journal of Chemical Physics*, 141:1–18, 2014.
[10] A. Wang and M. H. Müser. Is there more than one stickiness criterion? *Friction*, pages 4–7, 2022.
[11] F. Bottiglione, G. Carbone, L. Mangialardi, and G. Mantriota. Leakage mechanism in flat seals. *Journal of Applied Physics*, 106, 2009.
[12] W. B. Dapp, A. Lücke, B. N. Persson, and M. H. Müser. Self-affine elastic contacts: Percolation and leakage. *Physical Review Letters*, 108:1–4, 2012.
[13] A. I. Vakis, V. A. Yastrebov, J. Scheibert, L. Nicola, D. Dini, C. Minfray, A. Almqvist, M. Paggi, S. Lee, G. Limbert, J. F. Molinari, G. Anciaux, R. Aghababaei, S. E. Restrepo, A. Papangelo, A. Cammarata, P. Nicolini, C. Putignano, G. Carbone, S. Stupkiewicz, J. Lengiewicz, G. Costagliola, F. Bosia, R. Guarino, N. M. Pugno, M. H. Müser, and M. Ciavarella. Modeling and simulation in tribology across scales: An overview. *Tribology International*, 125:169–199, 2018.
[14] N. M. Kinkaid, O. M. O'Reilly, and P. Papadopoulos. Automotive disc brake squeal. *Journal of Sound and Vibration*, 267:105–166, 2003.
[15] A. Schallamach. How does rubber slide? *Wear*, 17:301–312, 1971.
[16] M. Barquins. Sliding friction of rubber and schallamach waves - a review. *Materials Science and Engineering*, 73:45–63, 1985.
[17] M. Aldam, Y. Bar-Sinai, I. Svetlizky, E. A. Brener, J. Fineberg, and E. Bouchbinder. Frictional sliding without geometrical reflection symmetry. *Physical Review X*, 6, 2016.
[18] N. Menga, G. Carbone, and D. Dini. Do uniform tangential interfacial stresses enhance adhesion? *Journal of the Mechanics and Physics of Solids*, 112:145–156, 2018.
[19] N. Menga, G. Carbone, and D. Dini. Corrigendum to "do uniform tangential interfacial stresses enhance adhesion?" [journal of the mechanics and physics of solids 112 (2018) 145–156]. *Journal of the Mechanics and Physics of Solids*, 133:103744, 12 2019.
[20] O. Ben-David and J. Fineberg. Static friction coefficient is not a material constant. *Physical Review Letters*, 106:1–4, 2011.
[21] S. Maegawa, A. Suzuki, and K. Nakano. Precursors of global slip in a longitudinal line contact under non-uniform normal loading. *Tribology Letters*, 38:313–323, 2010.
[22] M. K. Salehani, N. Irani, and L. Nicola. Modeling adhesive contacts under mixed-mode loading. *Journal of the Mechanics and Physics of Solids*, 130:320–329, 2019.
[23] N. Menga. Rough frictional contact of elastic thin layers: The effect of geometrical coupling. *International Journal of Solids and Structures*, 164:212–220, 2019.
[24] N. Menga, G. Carbone, and D. Dini. Exploring the effect of geometric coupling on friction and energy dissipation in rough contacts of elastic and viscoelastic coatings. *Journal of the Mechanics and Physics of Solids*, 148:104273, 2021.
[25] V. L. Popov, M. Heß, and E. Willert. *Normal Contact Without Adhesion*. Springer, 2019.
[26] M. K. Salehani, J. S. v. Dokkum, N. Irani, and L. Nicola. On the load-area relation in rough adhesive contacts. *Tribology International*, 144:106099, 2020.
[27] Y. Zhou and M. H. Müser. Effect of structural parameters on the relative contact area for ideal, anisotropic, and correlated random roughness. *Front. Mech. Eng.*, 6, 2020.
[28] A. Wang and M. H. Müser. Percolation and reynolds flow in elastic contacts of isotropic and anisotropic, randomly rough surfaces. *Tribology Letters*, 69:1–11, 2021.
[29] C. Campañá and M. H. Müser. Practical green's function approach to the simulation of elastic semi-infinite solids. *Physical Review B - Condensed Matter and Materials Physics*, 74:1–15, 2006.
[30] S. Sukhomlinov and M. H. Müser. On the viscous dissipation caused by randomly rough indenters in smooth sliding motion. *Applied Surface Science Advances*, 6:100182, 2021.
[31] M. Scaraggi and D. Comingio. Rough contact mechanics for viscoelastic graded materials: The role of small-scale wavelengths on rubber friction. *International Journal of Solids and Structures*, 125:276–296, 2017.
[32] See Supplemental Material at http://link.aps.org/supplemental/10.1103/PhysRevLett.131.156201 for model and methods: 3D Green's tensor implementation for GFMD simulations; generation procedure for single-wavelength and randomly rough surface profiles;

models for normal repulsion and lateral friction between the contacting bodies. Observables: calculation of the in-plane stress tensor; measurement of the resulting macroscopic friction force; numerical solution of the Reynolds equation. Details on Hertzian contact properties: comparison between multiple different values for Poisson's ratios and thickness in terms of in-plane and out-of-plane displacements, velocities, stress components and locally dissipated power.

[33] W. B. Dapp and M. H. Müser. Contact mechanics of and reynolds flow through saddle points: On the coalescence of contact patches and the leakage rate through near-critical constrictions. *Europhysics Letters*, 109:44001, 2015.

[34] Data-coupling-2023-07. https://github.com/sintharic/Data-Coupling-2023-07. Accessed: 2023-07-14.

[35] J. Scheibert, A. Prevost, G. Debrégeas, E. Katzav, and M. Adda-Bedia. Stress field at a sliding frictional contact: Experiments and calculations. *Journal of the Mechanics and Physics of Solids*, 57:1921–1933, 2009.

[36] N. Menga, C. Putignano, G. Carbone, and G. P. Demelio. The sliding contact of a rigid wavy surface with a viscoelastic half-space. *Proceedings of the Royal Society A: Mathematical, Physical and Engineering Sciences*, 470, 2014.

[37] C. Putignano, N. Menga, L. Afferrante, and G. Carbone. Viscoelasticity induces anisotropy in contacts of rough solids. *Journal of the Mechanics and Physics of Solids*, 129:147–159, 2019.

[38] S. C. Hunter. The rolling contact of a rigid cylinder with a viscoelastic half space. *Journal of Applied Mechanics*, 28:611–617, 12 1961.

[39] N. Menga, L. Afferrante, and G. Carbone. Effect of thickness and boundary conditions on the behavior of viscoelastic layers in sliding contact with wavy profiles. *Journal of the Mechanics and Physics of Solids*, 95:517–529, 2016.

[40] G. M. Hamilton and L. E. Goodman. The stress field created by a circular sliding contact. *Journal of Applied Mechanics, Transactions ASME*, 33:371–376, 1964.

[41] G. M. Hamilton. Explicit equations for the stresses beneath a sliding spherical contact. *Proc. Inst. Mech. Eng. C: J. Mech. Eng. Sci.*, 197:53–59, 1983.

[42] K. L. Johnson. *Contact Mechanics*. Cambridge University Press, 1985.

[43] B. N. Persson and C. Yang. Theory of the leak-rate of seals. *Journal of Physics Condensed Matter*, 20, 2008.

[44] D. Huang, X. Yan, R. Larsson, and A. Almqvist. Leakage threshold of a saddle point. *Tribology Letters*, 71:1–12, 2023.

[45] W. B. Dapp and M. H. Müser. Fluid leakage near the percolation threshold. *Scientific Reports*, 6:1–8, 2016.

[46] R. H. Bentall and K. L. Johnson. An elastic strip in plane rolling contact. *International Journal of Mechanical Sciences*, 10:637–663, 8 1968.

[47] J. Lengiewicz, M. d. Souza, M. A. Lahmar, C. Courbon, D. Dalmas, S. Stupkiewicz, and J. Scheibert. Finite deformations govern the anisotropic shear-induced area reduction of soft elastic contacts. *Journal of the Mechanics and Physics of Solids*, 143:104056, 2020.

[48] K. Vorvolakos and M. K. Chaudhury. Kinetic friction of silicone rubbers. *Langmuir*, 19:6778–6787, 2003.

# Supplementary materials

## Model and method

Model and method are similar to those used in many previous studies using GFMD, most notably in the original work [29], which, however, relied on atomistic Green's functions of Cu [111] surfaces rather than those valid in the continuum limit and on interfacial potentials lacking explicit interfacial dissipation. GFMD is contained in a few open source packages, however, the features needed to perform the simulations presented in our manuscript are not publically available to the best of our knowledge.

In principle, GFMD is a boundary-value method, which solves Newton's equation of motion for the Fourier coefficients of the surface displacements $\tilde{\mathbf{u}}(\mathbf{q})$. Convergence to the desired elastic deformation state can be achieved quickly by assigning inertia to surface modes with well-designed dependencies on the wave vector $\mathbf{q}$ of a given mode [27]. Calculation of stresses or forces acting on the surface mesh elements requires the elastic Green's functions and the interactions with the counterbody to be known. These aspects as well as other model and method details are described next in separate sections.

Before going into details, we alert the reader to a change of notation. In the main manuscript, a position in the interface is denoted as $\mathbf{r} = (x, y)$ and the (default) sliding direction is parallel to $x$. We switch to index notation $\mathbf{r} = (r_1, r_2)$ in the more technical appendix and assume the (default) sliding direction to be parallel to the unit vector $\mathbf{e}_1$, while using Einstein summation convention. Thus, the (default) sliding velocity would be denoted as $\mathbf{v}_0 = v_0 \mathbf{e}_\alpha \delta_{\alpha 1}$ in our notation. Since Cartesian indices run from 1 through 3, the index 0 in $v_0$ is not a Cartesian index.

### Green's functions

In linear, continuum theory, the elastic properties of an isotropic medium are defined by its Young's modulus $E$ and the Poisson's ratio $\nu$. For a frictionless contact of a semi-infinite linear elastomer, the relevant modulus is the contact modulus $E^* = E/(1-\nu^2)$, which is kept constant (unity) throughout this paper. For isotropic and homogeneous elastomers of thickness $h$, whose surface is flat in the absence of external stress, the Fourier coefficients or transforms of stress and strain are related through

$$\tilde{\sigma}_{3\alpha}(\mathbf{q}) = qE^* \Phi_{\alpha\beta}(\mathbf{q}, \nu, h, \cos\gamma)\tilde{u}_\beta(\mathbf{q}), \quad (1)$$

where $\Phi_{\alpha\beta}(...) = \Phi^*_{\beta\alpha}(...)$ and the interface normal is parallel to $\mathbf{e}_3$. Moreover, $\gamma$ is the angle formed by $\mathbf{q} = (q_1, q_2)$ and the in-plane displacement vector $\mathbf{r} = (r_1, r_2)$.

To simplify the dependencies of $\Phi_{\alpha\beta}(...)$ on the orientation between $\mathbf{q}$ and $\mathbf{u}$, it is easiest to express them in a coordinate system, in which $\mathbf{q}$ points parallel to the $r_1$ axis. Note that the current $r_1$-axis is, in general, not aligned with the sliding velocity $v_0 \mathbf{e}_1$. Thus, for $\mathbf{q} = (q_1, 0)$, the coefficients become [23, 31]

$$\frac{\Phi_{11}(hq, \nu)}{(1-\nu)^2} = \frac{(3-4\nu)\sinh(2qh) - 2qh}{(3-4\nu)^2 \sinh^2(qh) - (qh)^2} \quad (2a)$$

$$\frac{\Phi_{13}(hq, \nu)}{(1-\nu)} = \frac{iq_x}{q} \frac{(3-4\nu)(1-2\nu)\sinh^2(qh) - (qh)^2}{(3-4\nu)^2 \sinh^2(qh) - (qh)^2} \quad (2b)$$

$$\frac{\Phi_{22}(hq, \nu)}{1-\nu} = \frac{1}{2\tanh(qh)} \quad (2c)$$

$$\frac{\Phi_{33}(hq, \nu)}{(1-\nu)^2} = \frac{(3-4\nu)\sinh(2qh) + 2qh}{(3-4\nu)^2 \sinh^2(qh) - (qh)^2} \quad (2d)$$

and $\Phi_{12}(...) = \Phi_{23}(...) = 0$. Except $\Phi_{13}$, which is purely imaginary up to isolated points where it vanishes, coefficients in Eq. (2) are real and positive. In this latter case, displacements and stresses are in phase. The purely imaginary nature of $\Phi_{13}$ implies a phase shift of $\pm\pi/2$. Their effect is represented graphically in Fig. 1 and can moreover be summarized as follows, where $\tilde{\mathbf{G}}$ is the inverse matrix of $\mathbf{\Phi}$:

| cause | coupling effect |
|---|---|
| $u_1(x) = \hat{u}\cos(qx)$ | $\Delta\sigma_3 = +\mathrm{Im}(\Phi_{13})\hat{u}\sin(qx)$ |
| $u_3(x) = \hat{u}\cos(qx)$ | $\Delta\sigma_1 = -\mathrm{Im}(\Phi_{13})\hat{u}\sin(qx)$ |
| $\sigma_1(x) = \hat{\sigma}\cos(qx)$ | $\Delta u_3 = -\mathrm{Im}(\tilde{G}_{31})\hat{\sigma}\sin(qx)$ |
| $\sigma_3(x) = \hat{\sigma}\cos(qx)$ | $\Delta u_1 = +\mathrm{Im}(\tilde{G}_{31})\hat{\sigma}\sin(qx)$ |

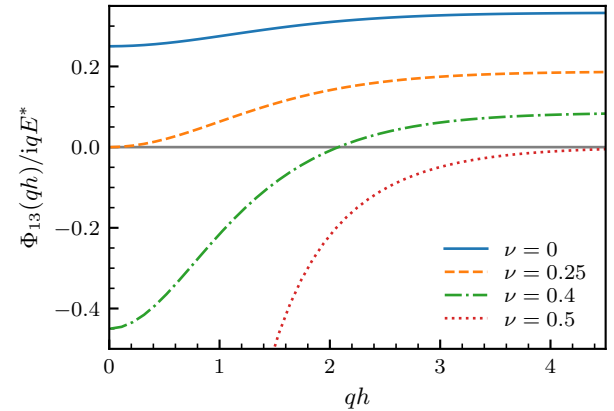

FIG. 1. Imaginary part of the (purely imaginary) coupling term $\Phi_{13}$ for different $\nu$ as a function of the product of wavevector $q$ and height $h$.

### Indenter geometries

The slider is assigned a height profile $h(\mathbf{r})$, which either is parabolic, consists of a square geometry or is ran-domly rough. In the first case, $h(r) = r^2/(2R_c)$, $R_c$ being the radius of curvature, while in the second case $h(\mathbf{r}) = h_0\{\cos(q_0 x) + \cos(q_0 y)\}/2$ with $q_0 = 2\pi/\lambda$ and $h_0 = \lambda/(2\pi)^2$, where $\lambda$ is the linear dimension of the periodically repeated simulation cell.

To generate the rough surface, the squared magnitude of a height Fourier coefficient, $|\tilde{h}(\mathbf{q})|^2$, is set to the height spectrum $C(q) \propto q^{-2-2H}$ for wave vectors $\mathbf{q}$ whose magnitude satisfies $2\pi/\lambda_l \le q \le 2\pi\lambda_s$, while all other $\tilde{h}(\mathbf{q}) = 0$. Here $H = 0.8$ is the Hurst roughness exponent, $\lambda_l = L/2$ is the long-wavelength cutoff, while $\lambda_s = \lambda_l/128$ the short wavelength cutoff. The phase of the Fourier coefficient of the height $\tilde{h}(\mathbf{q})$ is $2\pi$ times an independent, uniform random number $U(\mathbf{q}) \in (0,1)$ so that a height Fourier coefficient reads

$$\tilde{h}(\mathbf{q}) = \sqrt{C(q)} e^{i 2\pi U(\mathbf{q})}. \qquad (3)$$

The proportionality factor for $C(q)$ was chosen such that the root-mean square height is $h_{\text{RMS}} = 0.01\lambda_l$, which produces a root-mean-square gradient of $\bar{g} = 0.28$. Moreover, rough surface simulations were conducted with linear mesh resolutions of $\Delta x = \Delta y = \lambda_s/8$.

Results presented on stresses pertain only to one specific surface realization. Since stresses arise mostly in response to height gradients, stress distribution self-average quite quickly so that they do not change substantially from one random realization to the next, also because the system size was twice $\lambda_l$. Specifically, standard deviations of second moments are of order 5% of stress-tensor measures. Fluctuations are more significant for flow factors deduced from leakage calculations, because leakage currents are sensitive to long-wavelength undulations, in particular near the percolation threshold. This is why results for mean flow factors were averaged over eight roughness realizations.

*Interactions between elastomer and slider*

To obtain the quasi-static solution of the frictional contact, we take advantage of the spatio-temporal invariance, $x(t) = x(0) + v_0 t$, which results from the employed in-plane periodic boundary conditions and the elastomer initially being flat. Thus, in our simulations, the two surfaces are not explicitly moved with respect to each other. Instead, the in-plane velocity field of the elastic body (relative to the rigid indenter) is calculated as

$$\mathbf{v}^{\text{rel}}(\mathbf{r}) = v_0 \mathbf{e}_1 - \frac{d\mathbf{u}(\mathbf{r})}{dt} = v_0 \mathbf{e}_1 - \frac{\partial \mathbf{u}(\mathbf{r})}{\partial x} v_0. \qquad (4)$$

The Coulomb shear stress $\tau_C$ on a surface element is assumed to be antiparallel to $\mathbf{v}^{\text{rel}}$, but independent of its magnitude $v^{\text{rel}}$:

$$\tau_C(\mathbf{r}) = -\mu_c p_z \hat{\mathbf{v}}^{\text{rel}}(\mathbf{r}), \qquad (5)$$

where $\mu_c$ is the microscopic friction coefficient, and $\hat{\mathbf{v}}^{\text{rel}}(\mathbf{r})$ is the in-plane unit vector parallel to the relative velocity.

The interaction between slider and elastomer is modeled with a potential increasing quadratically with the local overlap, i.e., the interaction potential (before discretization) reads:

$$U_{\text{if}} = \int d^2 r \frac{k_{\text{if}}}{2} \{z(\mathbf{r}) - h(\mathbf{r})\}^2 \Theta\{z(\mathbf{r}) - h(\mathbf{r})\}, \qquad (6)$$

where $k_{\text{if}}$ is set close to the normal stiffness of the stiffest elastic mode, i.e., $k_{\text{if}} = 2E^*/\Delta x$, while $\Theta(...)$ denotes the Heaviside theta function. A short- but finite-range repulsion was chosen so that forces on mesh elements could be computed directly without having to deduce constraint forces first. The repulsion was made harmonic since this allows the time step to remain essentially as large as for a free-standing surface subjected to a simple time-dependent stress, which does not need to be determined self-consistently. Normal stresses can be deduced from first order (functional) derivatives of $U_{\text{if}}$ with respect to $z(\mathbf{r})$.

**Observables**

*Stress tensor calculation in rough interfaces*

To evaluate the stress-tensor at the interface, we use "Hooke's law"

$$\sigma_{\alpha\beta} = C_{\alpha\beta\gamma\delta} \varepsilon_{\gamma\delta},$$

with the symmetric strain tensor

$$\varepsilon_{\alpha\beta} = \frac{1}{2} \left( \frac{\partial u_\alpha}{\partial r_\beta} + \frac{\partial u_\beta}{\partial r_\alpha} \right).$$

Under the assumption of isotropy, the components of $C_{\alpha\beta\gamma\delta}$ can be expressed in dependence of only the known parameters of Young's modulus $E$ and Poisson's ratio $\nu$. Thus, we are provided with six equations for six known and six unknown variables. The known variables are the displacement derivatives $\partial u_\alpha/\partial r_{\beta \ne 3}$ and the stress-tensor elements $\sigma_{\alpha 3} = \sigma_{3\alpha}$, making it possible to solve the linear system of 6 equations for the 6 unknown quantities $\sigma_{\alpha \ne 3 \beta \ne 3}$ and $\partial u_\alpha/\partial r_3$.

*Friction force*

The kinetic friction $F_k$ can be deduced from the dissipated power via [19]

$$P_{\text{diss}} = \mu_c \int d^2 r \, p_z(\mathbf{r}) v_r(\mathbf{r}) \qquad (7)$$

through $F_\mathrm{k} = P_\mathrm{diss}/v_0$. Here, $p_z(\mathbf{r})$ is the normal pressure and $v_\mathrm{r}(\mathbf{r})$ the absolute in-plane velocity of a point on the surface of the elastomer relative to the slider and the integral taken over the contact area. The microscopic friction coefficient equals the macroscopic one, as long as the pressure profile is symmetric and deviations of the relative velocities from the center-of-mass velocity $v_0$ are antisymmetric, which is the case in the absence of coupling for a Hertzian tip and steady-state sliding.

The lateral force associated with a downhill-slope force is given by [36, 39]

$$\Delta F = -\int d^2 r \nabla h(\mathbf{r}) p_\mathrm{z}(\mathbf{r}). \quad (8)$$

We explicitly verified that our code produced the same corresponding total friction $F_\mathrm{k} = \mu_\mathrm{c} F_z + \Delta F$ via the downhill-slope argument as through a dissipation calculation.

*Flow factors*

The flow factors or leakage current are determined by solving the Reynolds thin-film equation, in which the local resistance to fluid flow scales with the inverse third power of the interfacial separation. To this end, we use a house-written code, which was developed for earlier work [33]. Mechanical stresses on the elastomer originating from fluid gradients are neglected.

Critical relative contact areas $a_\mathrm{c}^*$ were determined through nested intervals: The external pressure was iteratively adjusted and each time a flow calculation was performed until we found the exact pressure (and corresponding contact area), at which the flow factor in a given direction drops to 0.

### Details on Hertzian contact properties

Some of the discussions in the main text may benefit from line plots of data that was previously shown only as heat map. Fig. SM 2 summarizes the most important results for Hertzian contacts. Moreover, as alluded to in the main text and described in Eq. (7), changes in friction can also be rationalized by analyzing changes in the local dissipation. Without coupling, the normal stress is axisymmetric while the excess velocity field $\dot{\mathbf{u}}$ is antisymmetric with respect to a 180° rotation around the $z$-axis. With coupling, the only remaining plane of (anti-) symmetry is the $xz$ plane, which is most easily visible for the transverse velocity. In principle, an asymmetry is needed in the stress field and/or a symmetric component in the excess velocity in order for coupling to affect the overall heat production. In practice, it turns out that the coupling of the symmetric preexisting normal stress to the induced symmetric excess velocity is the dominating effect, which decreases friction for compressibility coupling but increases it for confinement coupling. The coupling of the induced antisymmetric stress field with the antisymmetric preexisting velocity field has the opposite effect but is of smaller magnitude.

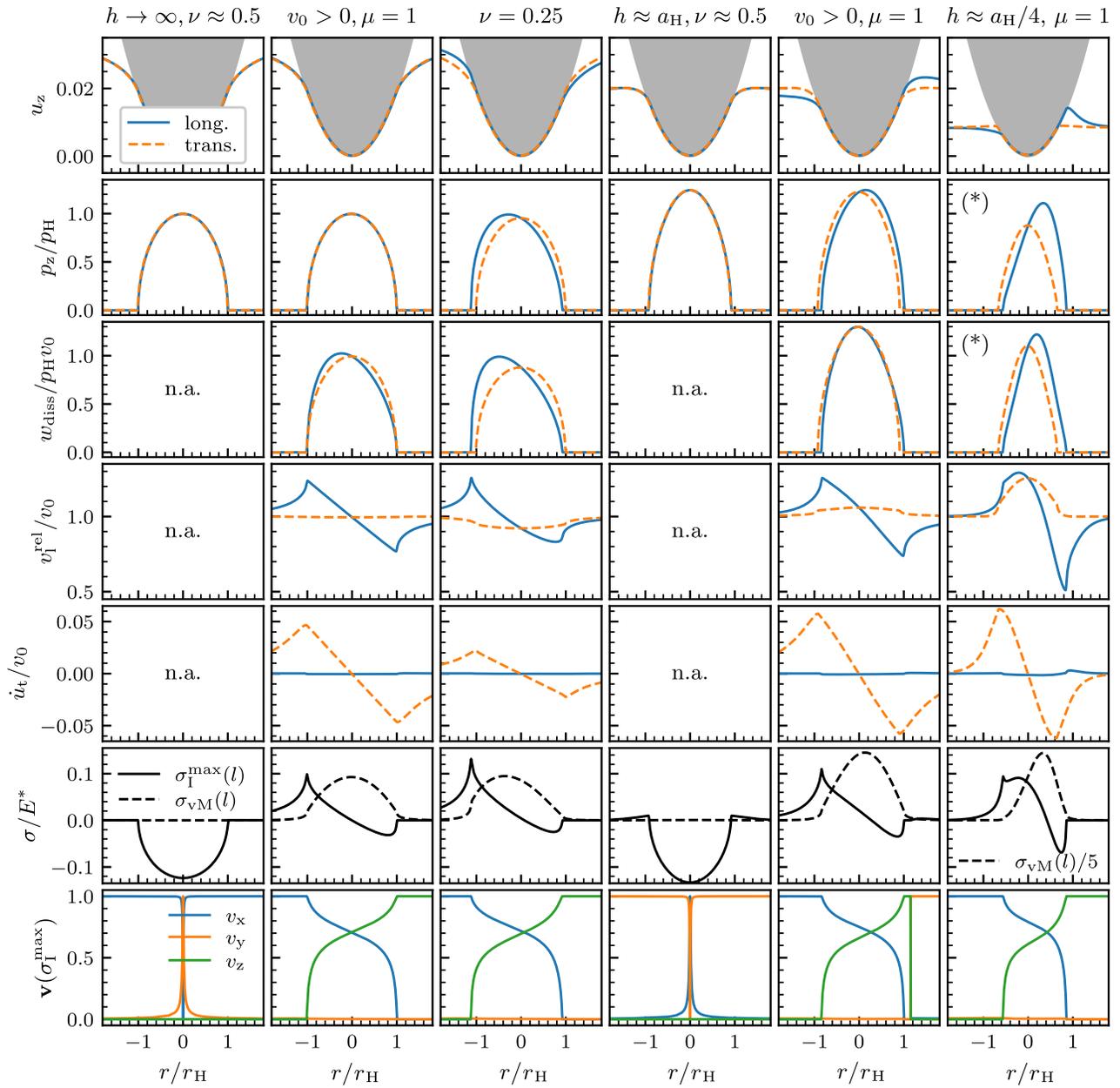

FIG. 2. Profile plots of contact properties for a rigid parabolic indenter in contact with an elastic material of varying properties. The microscopic friction coefficient was set 1, except for columns 1 and 4, which represent static cases without sliding or friction. Columns 1-3 represent semi-infinite elasomers, the first two cases being incompressible, whereas the third one has a Poisson's ratio of 0.25. Columns 4-6 show confined elastomers with finite thickness, where the last one is thinner than the first two. The seven rows represent the areal distributions of normal displacement, normal pressure, dissipated power, relative longitudinal velocity, transverse velocity, internal stresses and stress eigenvectors, respectively. Solid blue lines indicate a plot along the longitudinal (sliding) direction, dashed orange lines the transverse direction. In the last two rows, all properties are only shown in longitudinal (sliding) direction Data marked with "(*)" (last plot in the second and third row) was divided by 2.5 in order to match the scales of the other cases. Note that the plot on the far right in the sixth row also contains one rescaled data set. The last row contains the components of the normalized eigenvector belonging to the maximum eigenstress plotted in the row above. As introduced before, the indices l, t and z stand for longitudinal (parallel to sliding), transverse (perpendicular to sliding) and normal (to the surface), respectively.